\documentclass[aps,prl,twocolumn,showpacs,superscriptaddress]{revtex4-1}
\pdfoutput=1

\usepackage{graphicx}         
\usepackage{bm}               
\usepackage{amssymb}          
\usepackage{amsmath}          
\usepackage{appendix}		  

\usepackage{color}            

\begin{document}

\title{Multiple Ultralight Axionic Wave Dark Matter and Astronomical Structures}

\author{Hoang Nhan Luu}
\email{hnhanxiii@gmail.com}
\affiliation{Institute for Advanced Study and Department of Physics, Hong Kong University of Science and Technology, Hong Kong}

\author{S.-H. Henry Tye }
\email{iastye@ust.hk}
\affiliation{Institute for Advanced Study and Department of Physics, Hong Kong University of Science and Technology, Hong Kong}
\affiliation{Department of Physics, Cornell University, Ithaca, NY 14853, USA}

\author{Tom Broadhurst}
\email{tom.j.broadhurst@gmail.com}
\affiliation{Department of Theoretical Physics, University of the Basque Country UPV/EHU,
	E-48080 Bilbao, Spain}
\affiliation{Donostia International Physics Center (DIPC), 20018 Donostia-San Sebastian (Gipuzkoa)
	Spain.}
\affiliation{Ikerbasque, Basque Foundation for Science, E-48011 Bilbao, Spain}

\date{\today}

\begin{abstract}

An ultralight scalar boson with mass $m_1 \simeq 10^{-22}$ eV is gaining credence as a Dark Matter (DM) candidate that explains the dark cores of dwarf galaxies as soliton waves. Such a boson is naturally interpreted as an axion generic in String Theory, with multiple light axions predicted in this context. We examine the possibility of soliton structures over a wide range of scales, accounting for galaxy core masses and the common presence of nuclear star clusters. We present a diagnostic soliton core mass-radius plot that provides a global view, indicating the existence of an additional axion with mass $m_2\simeq 10^{-20}$ eV, with the possibility of a third axion with mass $m_3 \gtrsim 0.5 \times 10^{-18}$ eV. We also argue that the relative mass densities measured for these axions are consistent with their cosmological production via the mis-alignment mechanism.
\end{abstract}

\maketitle

\section{Introduction}

Dark Matter (DM) is well established from a wide set of astronomical evidence, including dynamical, lensing and CMB data. However, its nature is far from clear, requiring new physics beyond the known standard particle physics that describes only $\simeq 17\%$ of the cosmological mass density \cite{Ade2015xua,Cyburt2015mya}. It is understood that dark matter must be predominantly non-relativistic, to the earliest limits of observation, otherwise large-scale structure would be featureless on small scales. However, no evidence for heavy, non-relativistic weakly interacting massive particles (WIMPs) has been found, despite stringent laboratory searches. Alternatively, dark matter as a Bose-Einstein condensate also satisfies the non-relativistic requirement, as light bosons in the ground state behave as a condensate with high occupation number. The uncertainty principle means bosons cannot be confined within the de Broglie scale, thus avoiding the density cusp formation problem for WIMPs, as raised in the fuzzy dark matter scenario \cite{Hu2000ke} with an ultralight boson. \\

Ultralight bosonic dark matter has recently been successfully simulated for the first time, revealing an unforeseen rich wavelike substructure that may be termed ``wave dark matter" ($\psi$DM) \cite{Schive2014a, Schive2014b, Schive2016, Hui2016ltb, Niemeyer2019aqm}. A prominent solitonic wave at the base of every virialised potential is predicted by $\psi$DM, representing the ground state, where self gravity is matched by effective pressure from the uncertainty principle. The solitons found in the simulations have flat cored profiles that accurately match the known time independent solution of the Schr\"{o}dinger-Poisson equation \cite{Schive2014a,Schive2014b,Mocz2017wlg,Veltmaat2018}, for which the soliton mass scales inversely with its radius \cite{Guzman2006}. \\

Here we take seriously the generic String Theory prediction of multiple axion fields on astronomical scales. An axion mass has been derived in this context to be $m_1 \simeq 10^{-22}$ eV \cite{Hu2000ke,Schive2014a,DeMartino2018zkx}. Here, we will advocate a solitonic origin for the puzzling dynamically distinct, Nuclear Star Cluster (NSC) of $10^7 \text{ M}_\odot$ in the Milky Way that surrounds the central black hole on a scale of $\simeq 1$ pc. The inner density profile of this NSC is fitted by a dense soliton of dark matter corresponding to a heavier axion with mass $\simeq 10^{-20}$ eV. This inner soliton amounts to a small dark matter contribution in addition to the dominant soliton due to the $10^{-22}$ eV axion responsible for galaxy formation in this context, forming this ``soliton in soliton" structure within the Milky Way on scales of 100 pc and 1 pc for axions of $10^{-22}$ eV and $10^{-20}$ eV respectively. \\

\begin{figure*}[!ht]
	\centering
	\includegraphics[scale=0.7]{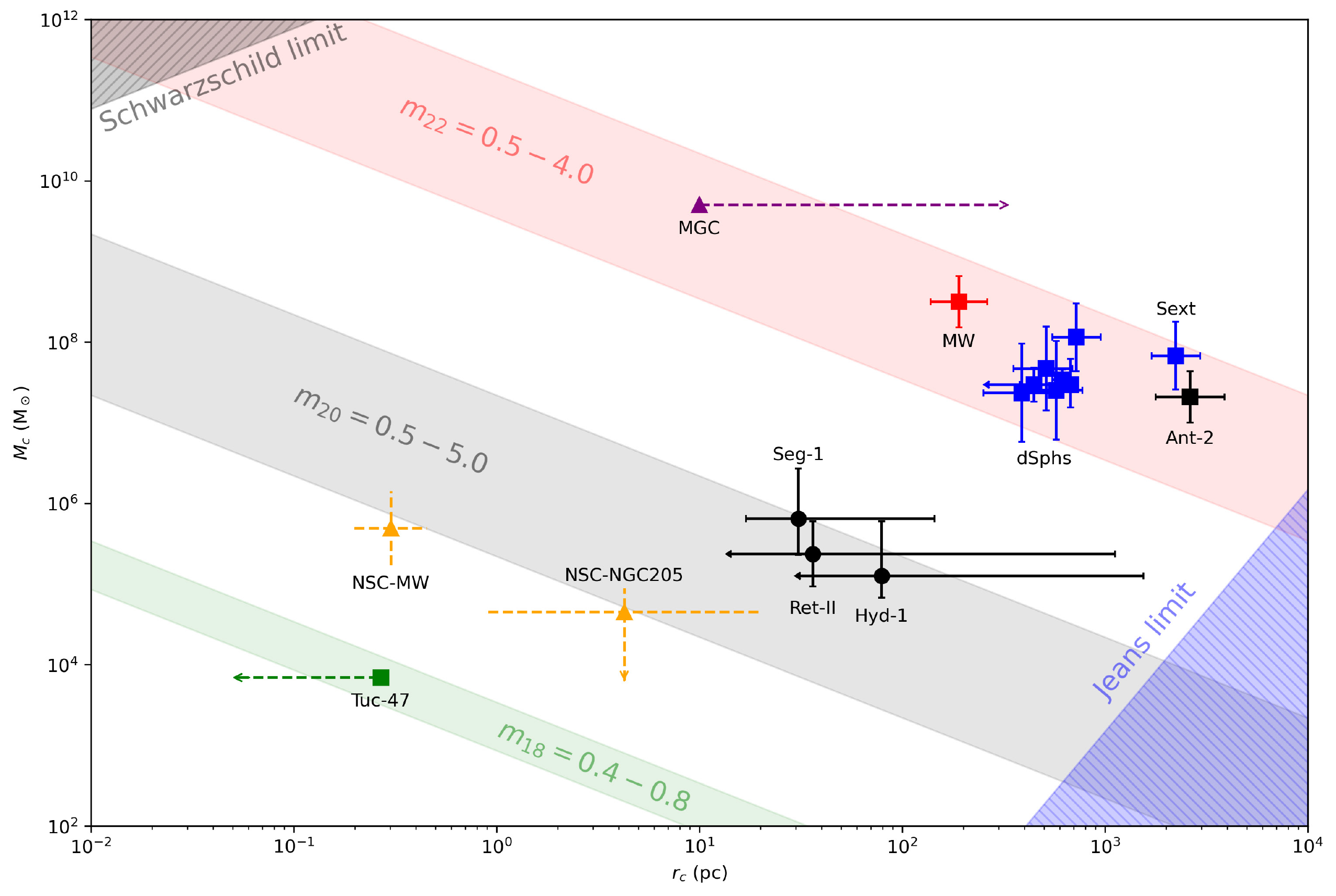}
	\caption{\textbf{The Soliton Core Mass-Radius Plot} illustrates the map of soliton core mass $M_c$ versus the core radius $r_c$ from cosmological structures at hierarchically different scales of the observational data points plotted together with theoretical prediction of solitons. The Milky Way (MW) and dSph galaxies data suggest an axion mass $m_1 \simeq 10^{-22}$ eV. The UFDs require another axion with $m_2 \simeq \times 10^{-20}$ eV and it is possible that NSCs are consistent with this more massive axion as indicated here tentatively for best studied examples, including the NSC in the Milky Way and NGC205. The dark core of the well studied globular cluster Tuc-47 data has also been suggested to harbour a soliton and the mass and scale of this corresponds to and axion of approximately $m_3 \gtrsim 0.5 \times 10^{-18}$ eV as shown lower left. Filled-colored bands are approximate mass ranges for the three possible axions masses
		that may be inferred as above for the current best data, spanning a wide range of astronomical objects that appear to dominated or to contain significant dark matter in excess of the visible stellar mass. The squares indicate objects for which axion mass constraints have been obtained previously and circles those obtained here.  The triangles show predicted scale sizes of solitons in order that they lie in given axion mass range. The Schwarzschild and the Jeans bounds are shown for reference. Dashed error bars indicate preliminary contraints for which substantial revision may be expected with a better determination of the scale of the associated dark matter.}
	\label{fig6}
\end{figure*}

The available data suggests that the primordial density of the lighter axion ($m_1$) is substantially bigger than that of the heavier ($m_2$) axion, even if the opposite should be true if they have equal number densities. Fortunately, this is consistent with cosmological considerations where the lighter scalar field is likely to have a higher density, if they are axions; that is, they are massless (due to shift symmetries) until non-perturbative dynamics generate the tiny masses. \\

In this paper, we first consider the solitonic properties of the multiple axion scenario, with a solution for ``nested" solitons corresponding to axions of different masses. We find that, to a reasonable approximation, solitons from different axions have negligible effect on each other. Based on this fact, we then form a global view by displaying all claimed relevant data in a plane of soliton core mass versus core radius, in Fig. \ref{fig6}; as shown, for a given axion mass $m_i$, there is a scaling solitonic solution, where the soliton core mass scales inverse to the core radius.
The main result of this paper, suggesting the existence of at least 2 distinct ultralight axions, with masses $m_1 \simeq 10^{-22}$ eV and $m_2 \simeq  10^{-20}$ eV, and the possibility of a third axion, $m_3 \gtrsim 0.5 \times 10^{-18}$ eV.  In addition to the data that we interpret as soliton-in-soliton phenomena, there are also data indicating the presence of single  $m_2$ solitons. Together, they provide support for the existence of $m_2$ in addition to $m_1$. 
Further focused testing of this unifying conclusion will clarify the extent to which these widely different astronomical structures can be understood as a manifestation of multi-ultralight axions. The data is also consistent with the axionic cosmology where the lighter axion is expected to have a higher primordial density. We conclude with predictions for new observations. \\

\section{Multiple Axions and Nested Solitons}

It is straightforward to extend the $\psi$DM formalism to that for the multiple axion case. At the first order perturbative theory and in the non-relativistic limit, we have the Schr$\ddot{\text{o}}$dinger-Poisson equations for N axion fields evolving on a Newtonian expanding background, which are coupled via $\Phi$,
\begin{align}
\begin{split} \label{eq1}
i\hbar&\dfrac{\partial \psi_1}{\partial t} = \left(-\dfrac{\hbar^2}{2m_1a^2}\nabla^2 + m_1\Phi\right) \psi_1, \\
&\vdots\\
i\hbar&\dfrac{\partial \psi_N}{\partial t} = \left(-\dfrac{\hbar^2}{2m_Na^2}\nabla^2 + m_N\Phi\right) \psi_N, \\
&\dfrac{\nabla^2\Phi}{4\pi Ga^2} = \left|\psi_1\right|^2 + ... + \left|\psi_N\right|^2 - \dfrac{3H^2}{8\pi G}.
\end{split}
\end{align}
Let us focus on the $N = 2$ case as this is relevant to this paper. For simplicity, we take into account the fact that the characteristic time for evolution of the system is short compared to the age of the universe, so $a$ becomes unity and $H$ vanishes. In addition, we also consider the system in symmetrically spherical coordinate and find the stationary solution expressed by $\psi_i (\textbf{x},t) = \psi_i(\textbf{x})e^{-iE_it/\hbar}$. Futhermore, to simplify physical constants, we rescale these quantities into the dimensionless variables \cite{Ruffini1969}
\begin{equation}\label{eq2}
\begin{aligned}
&r = \dfrac{\hbar^2}{2m^2 GM} \tilde{r},& & \psi_i = \left( \dfrac{2G^3m^6M^4}{\pi \hbar^6 M_i} \right)^{1/2}\tilde{\psi_i}, \\
& \Phi = \dfrac{2 G^2 M^2 m^2}{\hbar^2}\tilde{\Phi},& 
&E_i = \dfrac{2G^2M^2m^2}{\hbar^2}m_i\tilde{E_i}. 
\end{aligned}
\end{equation}
where $M_1, M_2$ are the total masses of the gravitational structure formed by each axion and $m$, $M$ are the scale parameters which could be determined in a specific system. $\Phi, E_1, E_2$ are one-particle potential and one-particle energy respectively, $\psi_1, \psi_2$ are one-particle wavefunctions which are normalized individually, $\int |\psi_i|^2 d^3\textbf{x} = 1$. This normalization of wavefunctions also implies $M_{tot} = \int \rho(\textbf{x}) d^3\textbf{x} = M_1 + M_2$, where $\rho(\textbf{x}) = M_1 |\psi_1|^2 + M_2 |\psi_2|^2$.  Finally, we obtain a system of scale-invariant equations
\begin{align}
\begin{split} \label{eq3}
&\dfrac{\partial^2 \tilde{\psi}_1}{\partial \tilde{r}^2} = -\dfrac{2}{\tilde{r}}\dfrac{\partial \tilde{\psi}_1}{\partial \tilde{r}} + \left(\dfrac{m_1}{m}\right)^2 (\tilde{\Phi} - \tilde{E}_1)\tilde{\psi}_1, \\
&\dfrac{\partial^2\tilde{\psi}_2}{\partial \tilde{r}^2} = -\dfrac{2}{\tilde{r}}\dfrac{\partial \tilde{\psi}_2}{\partial \tilde{r}} + \left(\dfrac{m_2}{m}\right)^2(\tilde{\Phi} - \tilde{E}_2)\tilde{\psi}_2, \\
&\dfrac{\partial^2\tilde{\Phi}}{\partial \tilde{r}^2} = -\dfrac{2}{\tilde{r}}\dfrac{\partial \tilde{\Phi}}{\partial \tilde{r}} + \left|\tilde{\psi}_1\right|^2 + \left|\tilde{\psi}_2\right|^2.
\end{split}
\end{align}
These equations can be solved numerically under some necessary constraints
\begin{align}\label{eq4}
\begin{split}
&\tilde{\psi}'_1(0) = \tilde{\psi}'_2(0) = \tilde{\Phi}(0) = 0\\
&\tilde{\psi}_1(\infty) = \tilde{\psi}_2(\infty) = 0,\\
&\int^\infty_0 |\tilde{\psi}_i|^2 \tilde{r}^2 d\tilde{r} = \dfrac{M_i}{M},
\end{split}
\end{align}
given the ratio $m_1/m_2$ and $M_1/M_2$, once we have solution in term of $\tilde{\psi}_i(\tilde{r})$ we can find the corresponding density profile at any scale by choosing appropriate value for $m$ and $M$. Because of the existing shift symmetry in these equations, we can choose an arbitrary value for the gravitational potential at the origin without changing the solution of $\psi(r)$. As an illustration, we approach the problem by initially setting the central values of wavefunctions $\tilde{\psi_1}(0)/\tilde{\psi_2}(0)$ and identify the normalization factors after the corresponding solution has found, as a result, if $\int^\infty_0 |\tilde{\psi}_i|^2 \tilde{r}^2 d\tilde{r} = \alpha_i$, the following density profile 
\begin{align}
\rho (r) = \dfrac{2M^4 m^6 G^3}{\pi \hbar^6}\left( \left| \tilde{\psi_1} \right|^2 + \left| \tilde{\psi_2} \right|^2 \right),
\end{align}
where $\tilde{\psi}_1, \tilde{\psi}_2$ are solution of \eqref{eq3}, describes a soliton with the total mass $M_{tot} = (\alpha_1 + \alpha_2)M$. \\

\begin{table}
	\centering
	\begin{tabular}{ |c|c|c|c|c|c| } 
		\hline
		$\tilde{\psi}_2(0)$ / $\tilde{\psi}_1(0)$ & $\tilde{E}_1$ & $\tilde{E}_2$ & $M_2/M_{tot}$\\
		\hline 
		0.5 & 1.68251566 & 0.69765202 & 2.7\% \\
		1.0 & 1.91670679 & 0.83895681 & 8.8\% \\
		2.0 & 2.59599377 & 1.24841522 & 23\% \\
		3.0 & 3.38136093 & 1.72370456 & 35\% \\
		\hline
	\end{tabular}
	\caption{Stationary double-solution solutions in terms of $\tilde{E}_1$ and $\tilde{E}_2$ considered for different central values of rescaled wavefunctions with $m_2 / m_1 = 3$. Notice that the accuracy of $\tilde{E}_1$ and $\tilde{E}_2$ is extremely sensitive to the behavior of the regular solution in this two-axion problem. Because the generic solution blows up due to the existence of a movable-pole at a finite radius $\tilde{r}_{sin}$, we need to impose the infinity boundary condition at some $\tilde{r}_{max} < \tilde{r}_{sin}$ and gradually move this pole to a larger radius. Hence, the solutions above are regular up to $r_{max} = 9$. $\tilde{E}_1$ and $\tilde{E}_2$ can also vary depending on the bins size of $\tilde{r}$ in Runge-Kutte algorithm, here taken as $\Delta \tilde{r} = 0.02$. The detailed numerical method we used to solve for these values will be presented in a future work (in preparation).}
	\label{soliton}
\end{table}

\begin{figure}[!ht]
	\centering
	\includegraphics[scale=0.35]{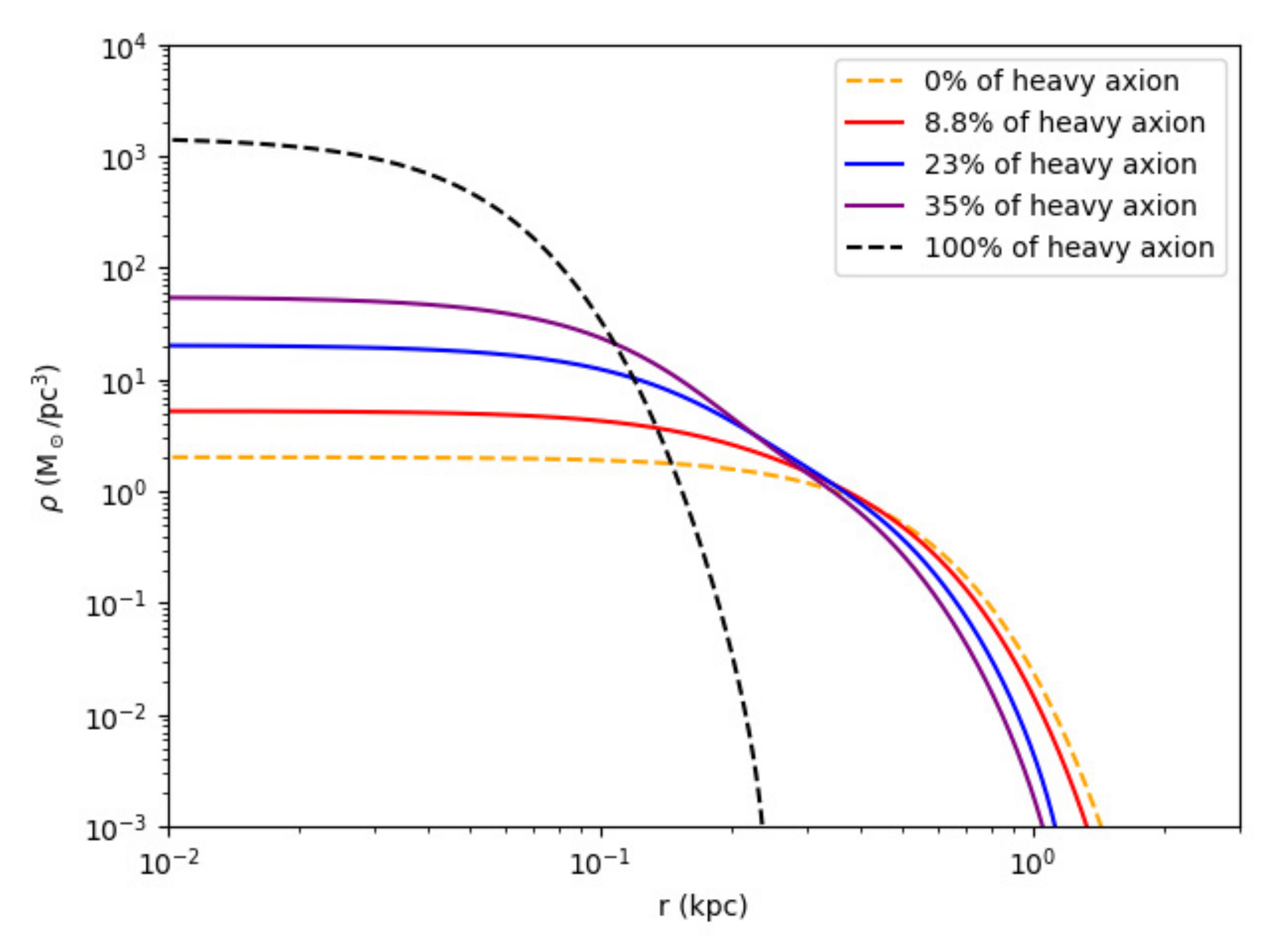}
	\caption{Soliton mass density profiles calculated in four cases of Table \ref{soliton}, with $M_{tot} = M_1 + M_2 = 10^{9} \text{ M}_\odot$ and $m_2 = 3 m_1 = 2.4 \times 10^{-22}$ eV, where $M_i$ is the total masses of individual soliton. We see that as the fraction of light axion increases, the two-axion profile approaches the single-axion one and vice versa.}
	\label{fig1}
\end{figure}

\begin{figure}[!h]
	\centering
	\includegraphics[scale=0.35]{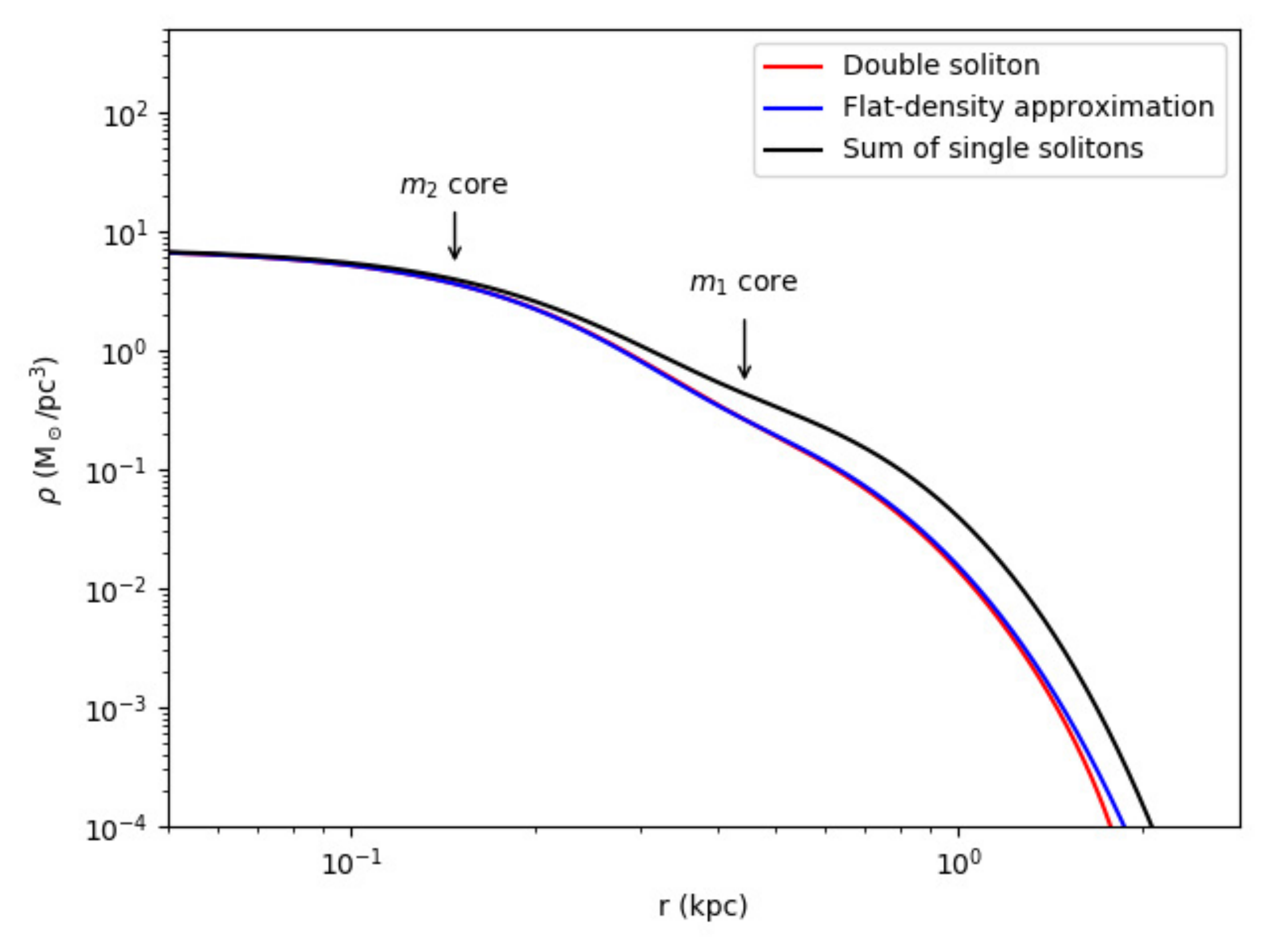}
	\includegraphics[scale=0.35]{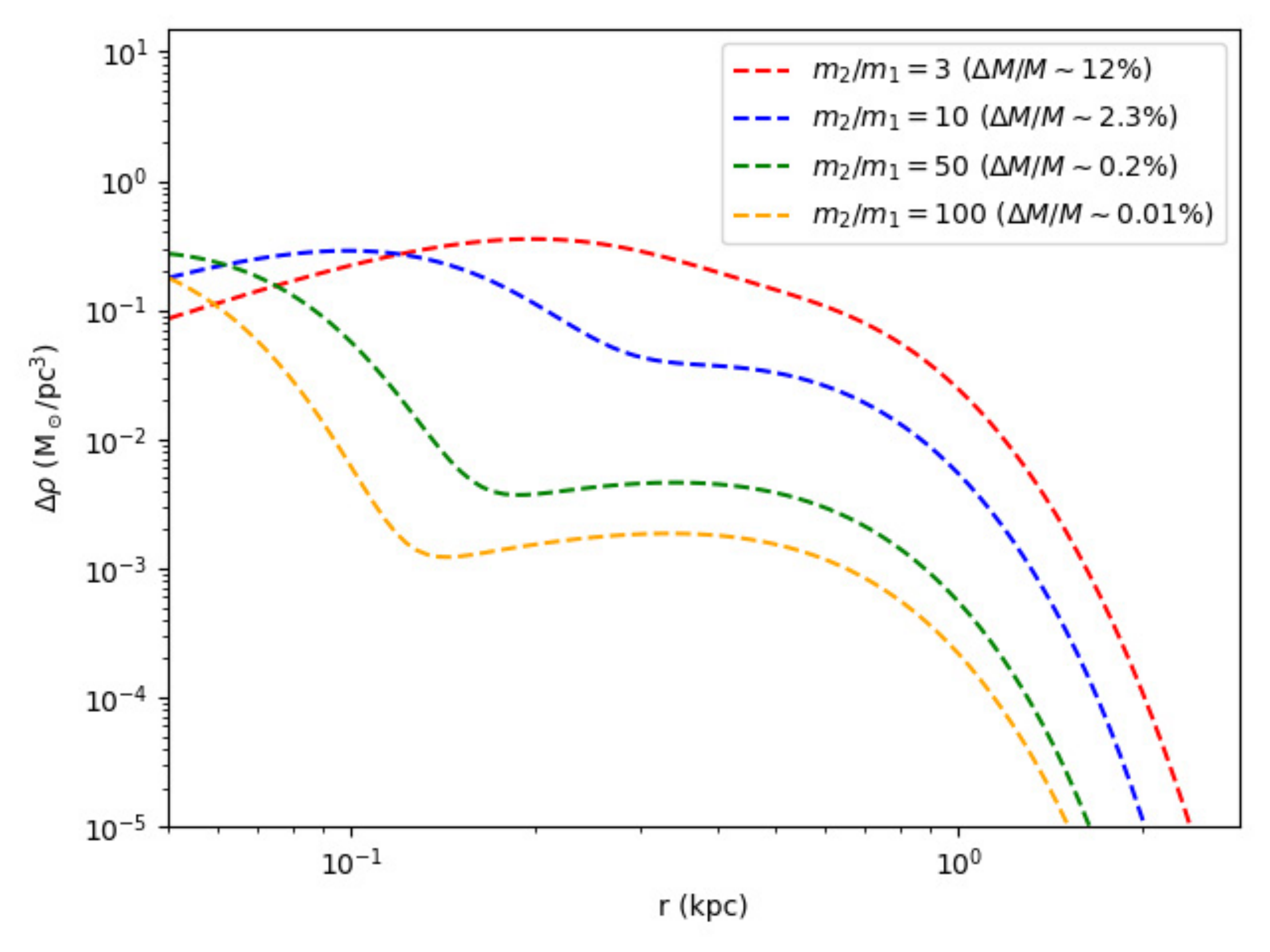}
	\caption{\textbf{Upper panel}: Nested soliton profile calculated with three different methods with the same central density in each case, $\rho_2(0) = 9\rho_1(0) \simeq 6.3 \text{ M}_\odot/\text{pc}^3$, and $m_2 = 3 m_1 = 8 \times 10^{-23}$ eV. The red-solid curve shows the numerical solution of the coupled two-axion system \eqref{eq1}, while the black-solid curve is obtained when simply adding two independent single-axion soliton profiles \eqref{eq7}, neglecting the coupling between them. The solid-blue curve (that almost overlaps with the solid red curve) is obtained numerically assuming the heavier axion is solely influenced by the relatively flat gravitational potential of the lighter axion. Details can be found in Appendix A. \textbf{Lower panel}: Dashed-colored curves show the density difference $\Delta \rho$ between the solid black and blue curves above for different $m_2/m_1>1$ fractions. The corresponding total mass deviation $\Delta M / M$ characterizes the correction in each case, where $M$ is the mass scale corresponding to each case, which is computed from the normalization condition.}
	\label{fig2}
\end{figure}

As an illustration, assume $m_1/m_2 = 3$ and $m^2 = m_1m_2$, a few solutions are listed in Table \ref{soliton}. Fig. \ref{fig1} compares these solutions with the universal single-axion profile suggested in \cite{Schive2014a}. Using the same setup with $m_2 = m_1/3 = 8 \times 10^{-23}$ eV and $\tilde{\psi}_1(0) = 3 \tilde{\psi}_2(0)$, we evaluate the magnitude of this mutual influence on the nested soliton profile in terms of the residual density deviation of the coupled case as a function of radius, as illustrated  in Fig. \ref{fig2}. The deviation from a simple sum of the two solitonic cores is approximately 12\%, as shown in the lower panel, and becomes much less than 1\% for our mass range of interest here, $m_2/m_1 \sim 100$. We conclude that soliton mass density profiles from different axions hardly influence each other's presence for the large axion hierarchies of relevance here (in Fig. \ref{fig6}), so we may accurately describe overlapping solitons from different axions with the simple sum of single mass density profiles \cite{Schive2014a},
\begin{align}
\rho_s(r) &= \rho_{s,1}(r) + \rho_{s,2}(r) \qquad \text{ where} \\
\rho_{s,i} (r) &\simeq \dfrac{1.9 \times 10^{10} (m_i/10^{-22}\text{ eV})^{-2}(r_{c,i}/\text{pc})^{-4}}{[1 + 9.1 \times 10^{-2}(r/r_{c,i})^2]^8} \text{ M}_\odot/\text{pc}^3, \label{eq7}
\end{align}
where the core radius $r_c$ is the radius at which the density is half the peak density, i.e., $\rho_s(r_c) = \rho_s(0)/2$. This numerically-fitted profile of $\rho_s(r)$ is well approximated up to $r \sim 3 r_c$. Each soliton follows its own individual core mass-core radius relation
\begin{align}
M_{c,i} &\simeq \dfrac{5.5 \times 10^{10}}{(m_i/10^{-22} \text{ eV})^2(r_{c,i}/\text{pc})} \text{ M}_\odot. \label{corem}
\end{align}
Here, the core mass $M_c$ is the mass enclosed within $r_c$ and the (total) soliton mass is $M_{sol} \sim 4 M_c$, with $r_{sol} \sim 5 r_c$. Since the independent solitons approximation in the soliton-in-soliton picture is best when we restrict ourselves to radii smaller than the core radii, we shall apply this result to within the core radii as much as possible.\\

\begin{table}
	\footnotesize
	\centering
	\begin{tabular}{ |c|c|c|c| }
		\hline
		$\text{Structure}$ & $m$ (eV) & $M_c (\text{M}_\odot)$ & $r_c$ (pc) \\
		\hline 
		Carina & $1.9^{+1.9}_{-0.7} \times 10^{-22}$ & $2.5^{+7.7}_{-1.9} \times 10^7$ & $575^{+201}_{-195}$ \\
		Draco & $8.1^{+4.8}_{-3.0} \times 10^{-23}$ & $1.1^{+1.9}_{-0.7} \times 10^8$ & $724^{+231}_{-175}$ \\
		Fornax & $1.6^{+0.2}_{-0.2} \times 10^{-22}$ & $3.4^{+1.2}_{-0.9} \times 10^7$ & $617^{+75}_{-67}$ \\
		Leo I & $1.5^{+1.2}_{-0.5} \times 10^{-22}$ & $4.7^{+10.8}_{-3.3} \times 10^7$ & $513^{+179}_{-158}$ \\
		Leo II & $5.1^{+7.5}_{-3.2} \times 10^{-22}$ & $1.0^{+7.7}_{-0.9} \times 10^7$ & $200^{+268}_{-85}$ \\
		Sculptor & $2.0^{+0.5}_{-0.3} \times 10^{-22}$ & $3.0^{+1.08}_{-1.1} \times 10^7$ & $447^{+66}_{-66}$ \\
		Sextans & $ 2.5 - 9.5 \times 10^{-23}$ & $2.6 - 17.8 \times 10^7$ & $1.5 - 3 \times 10^3$ \\
		Ursa Minor & $2.5^{+2.3}_{-1.0} \times 10^{-22}$ & $2.3^{+7.2}_{-1.8} \times 10^7$ & $389^{+200}_{-138}$ \\
		Segue 1 & $5.4^{+8.1}_{-4.0} \times 10^{-21}$ & $6.5^{+20.5}_{-4.2} \times 10^5$ & $31^{+114}_{-14}$ \\
		Reticulum II & $1.3^{+0.8}_{-1.0} \times 10^{-20}$ & $2.3^{+3.7}_{-1.4} \times 10^5$ & $36^{+1086}$ \\
		Carina II & $9.3^{+3.9}_{-5.7} \times 10^{-22}$ & $3.0^{+3.2}_{-1.4} \times 10^7$ & $ \lesssim 676 $ \\
		Hydrus 1 & $1.5^{+1.2} \times 10^{-20}$ & $1.3^{+4.8}_{-0.6} \times 10^5$ & $79^{+1469}$ \\
		Draco II* & $\sim 5.6 \times 10^{-22}$ & $\sim 1.7 \times 10^7$ & $\sim 105$ \\
		Triangulum II* & $ \sim 3.8 \times 10^{-22}$ & $\sim 2.4 \times 10^7$ & $\sim 160$  \\
		Eridanus II* & $\gtrsim 6.0 \times 10^{-20}$ & $\sim 1.5 \times 10^4$ & $\lesssim 13$ \\
		Antlia 2 &$ 0.6-1.4 \times 10^{-22}$ & $ 1.0 - 4.4 \times 10^7$ & $ 1.8-3.4 \times 10^3$ \\
		Milky Way & $ 0.6-1.4 \times 10^{-22}$ & $ 1.5-6.6 \times 10^8$ & $ 117-266$ \\
		NSC of MW & $ 2.2 - 10 \times 10^{-20}$ & $ 1.7 - 14.5 \times 10^5$ & $ 0.15-0.45$ \\
		NSC of NGC205 & $ \gtrsim 2.5 \times 10^{-20}$ & $ < 9.0 \times 10^4$ & $ 0.9 - 20$ \\
		MGC & $ \lesssim 10^{-22}$ & $\sim 5 \times 10^{9}$ &  $< 1000 $ \\
		Tucana 47 & $\gtrsim 5.4 \times 10^{-19}$ & $\sim 6.9 \times 10^3$ & $\lesssim 0.27$\\
		\hline
	\end{tabular}
	\caption{The axion mass $m$ (eV), the solitonic core mass $M_c$ ($M_\odot$) and core radius $r_c$ (pc) fitting with different cosmological structure including globular clusters, dwarf galaxies, elliptical galaxies, clusters of galaxies and nuclear star clusters. *The core mass values for Tri-II, Dra-II and Eri-II are inferred from different speculations on the total halo masses as in \cite{Calabrese2016hmp, Marsh2018zyw}: $2 \times 10^{10} \text{ M}_\odot$ and $1.2 \times 10^7 \text{ M}_\odot$ respectively.}
	\label{tab2}
\end{table}

\section{Solitonic Core Mass - Core Radius}

Now we can discuss the observational data for the $\psi$DM model in Fig. \ref{fig6} and Table \ref{tab2}. This soliton core mass-core radius plot provides evidence that more than one axion exists; it also allows us to make predictions (e.g., the size of the massive galaxy clusters). Note that the masses involved spans 7 orders of magnitude while the distance scale spans 5 orders of magnitude. The only theoretical parameters are the axion masses.\\

$\bullet$ \textbf{Dwarf Spheroidal Galaxies (dSphs)}: In the context of $\psi$DM, the phase space distribution of stellar velocities and positions have been fitted by \cite{Chen2017a,Schive2014a} with a Jeans analysis to the classical dwarf spheroidal galaxies, Carina, Draco, Fornax, Leo I, Sculptor, Ursa Minor, providing strong evidence for an ultralight axion of $\simeq 10^{-22}$ eV, that follows well the soliton mass - halo relation predicted by the $\psi$DM simulations\cite{Schive2016}. By assuming Gaussian distributions for the quoted values of axion masses and cores radius in Table 1 of \cite{Chen2017a}, the mass cores of these galaxies are calculated using the relation \eqref{corem}. In addition we include the Sextans dSph galaxy also analysed in the $\psi$DM context\cite{Broadhurst:2019fsl}.\\

$\bullet$ \textbf{Ultra-faint Dwarf Galaxies}: 
These faint objects, now classified as galaxies, are distinctly smaller and much less luminous than the classical dwarf spheroidals with the largest mass-to-light among MW satellites, dominated DM. Currently there is a paucity of accurate stellar velocity measurements for this class of galaxy so that it has not been possible to distinguish between a cored or cusped mass profile, making their origin quite uncertain. Here we apply our Jeans analysis using individual stars as a first step to obtain constraints for $\psi$DM for the four best studied examples \cite{Simon2019}: Segue 1, Reticulum II, Carina II, Hydrus 1, which are chosen because of their data availability and sufficient numbers of member stars (more than 10), as described in Appendix B. Our values for scale radius and mass of these four UFDs are  shown in black in Figure \ref{fig6}, where it is clear that although a relatively well defined values are obtained, they differ as a class from the dwarf spheroidals in terms of their location in this plane, requiring a significantly higher axion mass. Note, due to the large apparent size of Carina II and its velocity dispersion, it can be classified as a dSph galaxy, that we show falls nicely on the low axion mass track of Fig. \ref{fig6}. \\

We are aware of similar studies \cite{Calabrese2016hmp} for two other UFDs including Triangulum II, Draco II and \cite{Marsh2018zyw} for Eridanus II with methods that are tentative to infer axion masses. In particular, the authors of \cite{Calabrese2016hmp} suggests a large maximum halo mass of Tri-II and Dra-II about $2 \times 10^{10} \text{ M}_\odot$ based on the dynamics of MW satellites, while \cite{Marsh2018zyw} makes use of the stability of the old star cluster residing in the center of Eri-II. With stellar extent of only $< 20$ pc, Tri-II and Dra-II are more reasonably hosted by halos of $10^{8-9} \text{ M}_\odot$. The speculating disruption of the star cluster in Eri-II is also likely to be compensated by the tidal disruption effect caused by MW \cite{Schive2019rrw}. Substantially increased spectroscopy will allow in the future more accurately defined mass profiles for this class of small, dark matter dominated galaxies. \\

$\bullet$ \textbf{Ghostly Giant Galaxy}: The new discovery by the Gaia satellite of a nearly invisible, dark matter dominated galaxy, Antlia 2 (Ant-2), by tracking its star motions \cite{Torrealba2018}, poses a strong conflict with cold dark matter; but its very low mass and large size (more than double any known dwarf galaxy) fit very nicely with a $m_1$ axionic soliton core, as shown in Fig \ref{fig6} where it extends our predicted relation to lower mass,
closer the limiting Jeans scale. More detail can be found in Ref. \cite{Broadhurst:2019fsl}. \\

$\bullet$ \textbf{Massive Galaxy Cluster}: Ref. \cite{Chen2018a} reports a compact dark mass of $\sim 2 \times 10^{10} \text{ M}_\odot$ near the center of a massive galaxy cluster (MGC), from the radius of curvature of a small lensed structure in a well resolved background galaxy lensed through the center of a massive lensing cluster MACS1149 in recent deep Hubble Frontier Fields images. However, the size is undetermined, with a radius bounded by about a few hundred pc. This mass may be an offset black hole ejected from the central luminous galaxy or a compact soliton with mass of $10^{10} \text{ M}_\odot$ that is expected at the bottom of the potential of a massive cluster of $10^{15} \text{ M}_\odot$. Assuming that this is due to the $m_1$ axion, we predict an extremely compact soliton with core radius $\simeq 10$ pc, corresponding to the smaller de-Broglie wavelength of a massive cluster. \\

$\bullet$ \textbf{Nested soliton in Milky Way and Central Nuclear Star Cluster}: Using a scaling relation derived from the simulations between the mass of soliton core and its host virial mass, $M_c \propto M_h^{1/3}$ \cite{Schive2014a,Schive2014b,Schive2016,Veltmaat2018}, Ref. \cite{DeMartino2018zkx} matches the missing mass recently found within the central $\simeq 100$ pc of Milky Way \cite{Portail2016,Zoccali2014} with a soliton of $\sim 10^9 \text{ M}_\odot$ through Jeans analysis, corresponds to an axion with mass $\simeq 8 \times 10^{-23}$ eV. This massive concentrated soliton explains well the projected radial enhancement of bulge star velocity dispersion peaking at $\simeq 130$ km/s, that is $50$km/s above the general bulge level of $\simeq 80$ km/s. \\

\begin{figure}[!ht]
	\centering
	\includegraphics[scale=0.35]{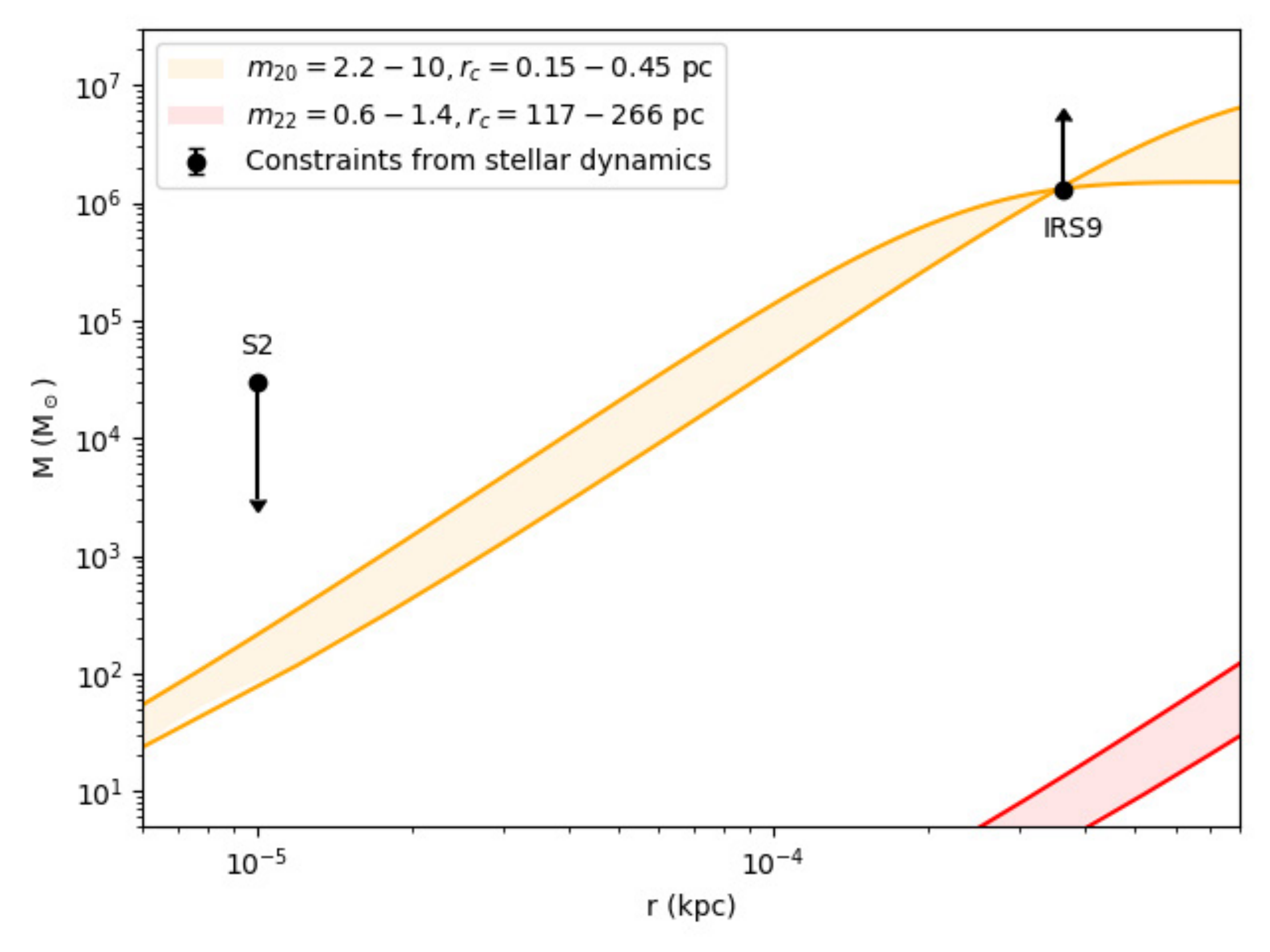}
	\caption{Fitting the soliton profile with the enclosed mass of the Nuclear Star Cluster in the Milky Way, using the motion of the star IRS9 at $r \sim 0.365$ pc from the central black hole \cite{Reid2006cu}. Here the enclosed mass $M(r)$ inside radius $r$ is computed by integrating the soliton profile \eqref{eq7}, after subtracting a black hole mass of $\sim 4 \times 10^6 \text{ M}_\odot$ and stars at the center of NSC. As a reference for comparison, we also highlight the mass range (the blue band) for the mass profile of solitons corresponding to the lighter axion $m_1$ which have a much wider core radius and relatively low central density \cite{DeMartino2018zkx}. Note, the accurate orbit of the closest orbiting star S2 provides a useful upper bound on any extended matter additional to the black hole \cite{Abuter2018drb}.}
	\label{fig5}
\end{figure}

The origin of the dynamically distinct nuclear star cluster present in the Milky Way (and in most other galaxies) remains unclear, in particular the presence of a core profile on a scale of $\simeq 1$ pc \cite{Buchholz2009} in the old star population, for which a old stellar cusp is firmly predicted but not evident \cite{Merritt2010}. The presence of excess unseen matter of $\simeq 10^6 \text{ M}_\odot$ may be implied on a scale of $\simeq 0.4$ pc, revealed by the high velocity orbit of the maser star IRS9, and excessive proper motions of other stars at this radius \cite{Schodel2018}, that imply an extended mass, additional to the central black hole on a parsec scale. Whether this mass can be accounted for by stars of the surrounding NSC is unclear, requiring a better understanding of the stellar mass function in this region. An upper limit to the mass of the DM is set by the dynamics of the stellar motion in the NSC which implies a	total mass of $3\times 10^7 \text{ M}_\odot$ on a scale of 3-5 pc radius of the NSC. Most of this dynamical based mass is thought to be stellar, though the uncertain choice of initial stellar mass function (IMF) means that as much as half this mass may not be stellar in the case of a Chabrier IMF rather than the Salpeter form \cite{Schodel2018, Alexander2005}. Under the conditions pointed above, we derive a smaller soliton by matching the enclosed mass at the location of star IRS9 while also satisfying an upper bound closer to the center from the orbit of star S2 \cite{Abuter2018drb}. This ``inner" soliton implies a heavier axion, $m_2 \simeq 10^{-20}$ eV, than the lighter axion $m_1$ responsible for the ``outer" soliton associated with bulge star dynamics of the Milky Way described above and shown in Fig. \ref{fig5}. The solution of an intermediate black hole on top of the outer soliton \cite{Brax:2019npi} for this structure is viable but less likely providing that the two bounds are both satisfied.\\

$\bullet$ \textbf{Early-type Galactic Nuclei of NGC 205}: Using high resolution data from Hubble Space Telescope, Ref. \cite{Nguyen2018} shows statistical evidence for the lack of a central black hole inside the dwarf elliptical (dE) NGC 205, where its nuclear star cluster resides. The decline of stellar mean velocities towards the center might imply the existence of a dark soliton of $\sim \times 10^5 \text{ M}_\odot$. However, uncertainties in the measured photometric and dynamical masses of that nuclear star cluster does not rule out the possibility of no soliton there, leaving only an upper bound of core mass. We assume reasonably that the soliton radius to be of order of the star cluster's size and with the lower bound is not too small to avoid having a point-like gravitational potential that is excluded by the
declining central velocity dispersion \cite{Nguyen2018}. Note that the extent of any solitonic DM profile is very unclear in the data, but can be reasonably expected extend beyond the stellar profile as it would have acted as a seed for gas attraction, so that stars that subsequently form from the cooling of this gas should not be expected to orbit beyond the soliton radius but to be generally of a smaller scale. This means the total mass of any such dark solitonic component may be substantially higher - nearly an order of magnitude may not be surprising if core radius is twice as large as the scale length of the stars. Hence we treat our constraint here cautiously, as indicated by the dashed error bars in Fig. \ref{fig6} and we regard these values as underestimates of the radius and mass which we may not be surprised to be revised towards a lighter axion mass than currently indicated in Fig. \ref{fig6}. \\

$\bullet$ \textbf{Globular Cluster 47 Tuc}: Compact dark mass has been reported in one of the best studied Globular Cluster within the Milky Way of $\simeq 3 \times 10^3 \text{ M}_\odot$  and naturally interpreted as long anticipated intermediate mass black holes \cite{denBrok2013, Kiziltan2017}. However, support for a point mass is shown to be lacking in the case of 47 Tuc in a new high resolution stellar proper motion study \cite{Mann2018xkm} that prefers an extended excess of binary stars and stellar mass black holes combined on a scale of 0.1 pc. A soliton explanation has been advocated for these dark excesses of 47 Tuc \cite{Emami2018rxq} corresponding to an axion of mass $\simeq 0.5 \times 10^{-18}$ eV, to account for the central dark core mass with an upper bound for the soliton core radius $\sim  0.03$ pc. For a smaller $r_c$, the axion becomes heavier. More data is needed to definitively test this tentative axion-soliton explanation. \\

\section{Cosmological evolution of axion relic abundance}

Based on the astrophysical data, we expect the $m_1$ axion contributes substantially more than the $m_2$ axion to the primordial dark matter density, while the contribution from the $m_3$ axion, if present, is negligible. This can be estimated roughly from the typical size needed to host a nested local structure. For the one in Milky Way, the mean density over the same volume is given by
\begin{align}\label{ratio1}
    \dfrac{\bar{\rho}_{MW}}{\bar{\rho}_{NSC}} = \dfrac{M_{c, MW}}{M_{c, NSC}} \simeq 0.5 \times 10^3.
\end{align}
It is reasonable to expect that the relative primordial densities of the $m_1$ axion and the $m_2$ axion is not far from this value,
\begin{align}\label{ratio2}
    \frac{\bar{\rho}_{1}}{\bar{\rho}_{2}} \gg 1.
\end{align}
If the primordial density $\bar{\rho}_2$ of the $m_2$ axion is much bigger, one expects many more single $m_2$ solitons than observed.
Qualitatively, this is consistent with the axion physics in cosmology. \\

In the early universe, an axion potential is flat as the axion $\phi$ is massless due to the presence of the shift symmetry, $m_a \sim 0$. So an axion potential contributes an energy density like the vacuum energy whose density stays more or less constant as the universe expands, until a mass is generated dynamically. This happens when the Hubble parameter is close to its mass, i.e., $3H \sim m_a$, leading to a potential for the axion,
\begin{align} \label{axionpot}
    V(\phi)\simeq m_a^2f_a^2 \left[ 1 - \cos (\phi/f_a) \right] \sim m_a^2\phi^2/2
\end{align}
where $f_a \gg m_a$ is the so called decay constant. Since there is no reason that the expectation value of $\phi$ happens to sit precisely at the bottom of the potential, it starts at $\phi_{ini} \sim f_a\theta_{ini}$ and roll down towards the bottom and begin to oscillate coherently around the bottom, which has the (simplified) equation of motion
\begin{align} \label{axionEoM}
    \ddot{\phi} + 3 H \phi + m^2_a \phi = 0
\end{align}
This is a damped simple harmonic oscillator.
When coherent oscillation begins, we can average over a period to obtain
the average energy density $\rho$ (up to a constant factor)
\begin{align}
    \bar{\rho} \simeq [\dot{\phi}^2 + m^2_a\phi^2]/2
\end{align}
which leads to (recall $H=\dot a/a$ where $a$ is the cosmic scale factor)
\begin{align}
    \dot{\bar{\rho}} = \left( -3H\right)\bar{\rho} \to {\bar{\rho}}\propto {1/a^3}
\end{align}
so it behaves like matter density. This is the mis-alignment mechanism. \\

Now, dynamics clearly suggests that the conversion from vacuum energy density to dark matter density at $a_{osc,j}$ is earlier for a heavier axion, so the $m_2$ axion matter density is diluted earlier than the $m_1$ axion density, ending with a lower $m_2$ axion energy density. In terms of the initial densities $\rho_{ini,j}$, we have
\begin{align} \label{MW_ratio}
    \dfrac{\bar{\rho}_{1}}{\bar{\rho}_{2}} = \dfrac{\rho_{ini, 1}}{\rho_{ini, 2}} \left( \dfrac{a_{osc, 2}}{a_{osc, 1}} \right)^3,
\end{align}
where $a_{osc, j}$ can be found by the condition $3 H(a_{osc, j}) = m_j$. Because axion fields with the mass range of our interest started oscillating well before matter-radiation equality, let us consider a radiation-dominated universe. Assume that the initial densities of two axion fields are comparable (see below), an exact solution can be solved in Fig. \ref{misalign}. On average, their relic abundance ratio in late universe is roughly
\begin{align}\label{MW_ratio2}
    \dfrac{\bar{\rho}_1}{\bar{\rho}_2} \simeq \dfrac{\rho_{ini, 1}}{\rho_{ini, 2}}\left( \dfrac{m_1}{m_2} \right)^{-3/2} \simeq \dfrac{\rho_{ini, 1}}{\rho_{ini, 2}}\times 10^3.
\end{align}
Let us consider the normalization for $\rho_{ini, j}$ for the $j$ axion,
$$\rho_{ini, j} = f_j^2 m_j^2\theta_{ini, j}$$
where the initial angle $\theta_{ini, j}$ takes a random value of order $\theta_{ini, j} \lesssim {\cal O}(1)$ radian. A priori, the $f_j$ are unrelated. However, if the non-perturbative effect of the two axions are due to the same dynamics, then it is reasonable to take $f_1^2m_1^2= f_2^2m_2^2$ (recall the analogous situation for the QCD axion, where it is related to the strong interaction pion via the quark condensate, $f_{\pi}^2m_{\pi}^2 = f_a^2m_a^2$.)
If so, then 
\begin{align}\label{ratio3}
{\rho_{ini, 1}}/{\rho_{ini, 2}} \sim 1 \quad \rightarrow \quad
{\bar{\rho}_1}/{\bar{\rho}_2} \sim 10^3
\end{align}
 This ratio is consistent with \eqref{ratio2} despite being inferred from two different epochs, showing that our two-axion picture is  consistent throughout the history of the Universe. 
 
 In fact, the ratio ${\rho_{ini, 1}}/{\rho_{ini, 2}}$ can be easily many orders of magnitude away from unity if their masses have different dynamical origins. That it is close to unity when we compare (\ref{ratio1}), (\ref{MW_ratio2}) and (\ref{ratio3}) hints that these two axion masses are generated by the same underlying non-perturbative dynamics. Applying this to the $m_3$ axion implies another 2-3 orders of magnitude lower in the relic density for the $m_3$ axion, which is negligible compared to the dominant $m_1$ axion density, as the observational data indicates.

\begin{figure}[!ht]
	\centering
	\includegraphics[scale=0.35]{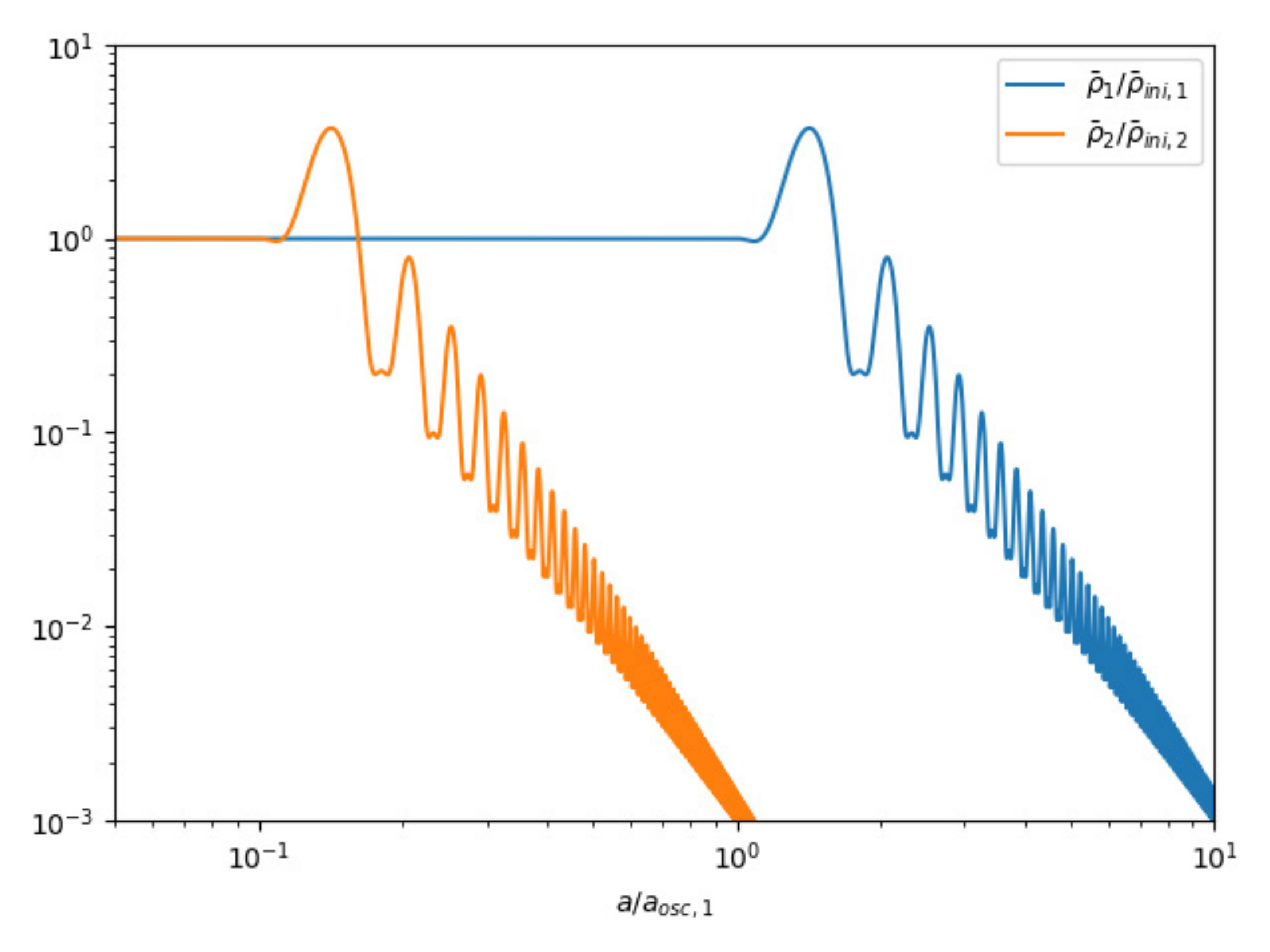}
	\caption{The mis-alignment density evolution of the two axion fields with mass $m_1$ and $m_2$. In the radiation-dominated universe, the equation \eqref{axionEoM} is transformed to $d^2 y / dx^2 + (2 / x) dy / dx + 9 x^2 y = 0$, where $y = \phi / \phi_{ini}$ and $x = a / a_{osc} = a / \sqrt{3 H_0 / m_a}$; then it can be solved numerically with the initial conditions $y(x = 1) = 1$, $y'(x = 1) = 0$. The normalized density is given by $\rho / \rho_{ini} = \left[(1 / 3x) dy / dx\right]^2 + y^2$. The fields behave as dark matter with $\bar{\rho} \sim a^{-3}$ when averaging over its oscillating period, which is roughly $m_a^{-1}$, as mentioned in the text.}
	\label{misalign}
\end{figure}

\section{Discussions and Conclusion}

We have focused on the soliton feature as a central prediction of the wave dark matter picture and generalized to the case of multiple axions, motivated by String Theory, deriving a stable time-independent, joint ground  state solution. We conclude that when the mass scales of two axions are spaced by more than a factor of a few in mass, the combined solution asymptotes to a nested pair of concentric solitons that are dynamically independent and this allows us to make clear comparisons with the observed structure at the center of our galaxy and other well resolved galaxies where a dense nuclear star cluster is commonly present. \\

We have shown that the lightest axion of $m_1\simeq \times 10^{-22}$ eV implied by both the Fornax Galaxy \cite{Schive2014a,Schive2014b,Schive2016} and the central Milky Way \cite{DeMartino2018zkx}, can also help interpret the massive compact lens of $10^{10}\text{ M}_\odot$ uncovered recently near the center of a massive galaxy cluster \cite{Chen2018a}.  \\

Our new Jeans-based dynamical analysis indicates strongly that second more massive axion of $m_2 \simeq 10^{-20}$ eV is required to explain dark matter dominated class of newly discovered "Ultra Faint Galaxies" (UFDs) found within the Milky Way, because of the  relatively small scale and high density of the dark matter compared to all other galaxies, including the classical dwarf Spheroidal class.  This second axion, together with the very small orbits of these UFDs within the Milky Way may imply these galaxy-like objects are more like nuclear star clusters that may have survived extensive tidal destruction of the host galaxy. This picture is tentatively supported by the possible presence of an excess of dark matter in the two best studied nuclear star clusters, that we have analysed here, at the center of the Milky way and at the center of the nearby NGC205 dwarf galaxy, that we have shown indicate a similar axion mass like the UFDs. \\

Finally, and more tentatively a compact inner mass of $\simeq 3 \times 10^3 \text{ M}_\odot$ found in the globular cluster 47 Tuc data may imply a further axion mass of $m_3 \gtrsim 0.5 \times 10^{-18}$ eV.  We have combined this information in a single core mass-radius plot (Fig. \ref{fig6}) providing a comprehensive ``birds-eye" view of the axion role in structure formation, concluding that these distinctive astronomical structures require more than one axion in the wave dark matter context.  \\

A definitive test for the presence of a nested soliton structure may be sought using pulsar timing residuals imprinted on millisecond pulsars detected at the Galactic Center \cite{Khmelnitsky2013lxt,DeMartino2017qsa}. Many thousands of pulsars are expected \cite{Calore2015bsx}, and can account for the GeV gamma-ray excess \cite{Abazajian2011,Abazajian2014}, which are being searched with some success \cite{Deneva2008rz}. Detection of such pulsars is a major goal of the SKA, including within the NSC region, given the high density star formation history in this cluster. Such NSC pulsars located within the nested solitons may show  distinctive multi-frequency timing residuals on the respective Compton time scales of these independently oscillating scalar fields, of a few hours and a few months corresponding to $10^{-20}$ eV and $10^{-22}$ eV respectively, providing a unique soliton signature as a definitive solution to the long standing Dark Matter puzzle. \\

\section*{Acknowledgments}

We thank Tzihong Chiueh and George Smoot for valuable discussions. SHHT is supported by the AOE grant AoE/P-404/18-6 issued by the
Research Grants Council (RGC) of the Government of the Hong Kong SAR China.

\begin{appendices}
	\section{Flat-density approximation}
	
	Although we have numerically solved the Schrodinger-Poisson equations of nested system in case the masses of two axions are not quite different, $m_2/m_1 = 3$, it is not practical to exactly solve these equations when this mass difference is significant because we have to fine-tune two eigenvalues $\tilde{E}_1$ and $\tilde{E}_2$ to maintain the stability of the wavefunction of the heavier axion. This technical issue is generic in finding numerical solution with multiple boundary conditions \cite{Moroz:1998dh}. On the other hand, we know the heavy-axion and light-axion wavefunction become wider and narrower respectively as either mass or central density difference increases. This fact implies that in the limit
	\begin{align}
	m_2 \gg m_1 \quad \text{and} \quad \tilde{\psi_2}(0) \gg \tilde{\psi_1}(0),
	\end{align}
	the heavy-axion wavefunction approximately evolves on a flat density distribution background of the light axion. Therefore, we can separately write the equations governing the evolution of two axions as follow	
	\begin{align*}
	&\dfrac{\partial^2 \tilde{\psi}_2}{\partial \tilde{r}^2} = -\dfrac{2}{\tilde{r}}\dfrac{\partial \tilde{\psi}_2}{\partial \tilde{r}} + (\hat{\Phi} - \tilde{E}_2)\tilde{\psi}_2, \\
	&\dfrac{\partial^2\hat{\Phi}}{\partial \tilde{r}^2} = -\dfrac{2}{\tilde{r}}\dfrac{\partial \hat{\Phi}}{\partial \tilde{r}} + \left|\tilde{\psi}_2\right|^2 + \left|\tilde{\psi}_1(0)\right|^2,
	\end{align*}
	and
	\begin{align*}
	&\dfrac{\partial^2\tilde{\psi}_1}{\partial \tilde{r}^2} = -\dfrac{2}{\tilde{r}}\dfrac{\partial \tilde{\psi}_1}{\partial \tilde{r}} + (\tilde{\Phi} - \tilde{E}_1)\tilde{\psi}_1, \\
	&\dfrac{\partial^2\tilde{\Phi}}{\partial \tilde{r}^2} = -\dfrac{2}{\tilde{r}}\dfrac{\partial \tilde{\Phi}}{\partial \tilde{r}} + \left|\tilde{\psi}_{2\text{ (solved)}}\right|^2 + \left|\tilde{\psi}_1\right|^2.
	\end{align*}
	They simply return to the familiar single-axion system of equations with an external source of flat density. Here, we use $\hat{\Phi}$ and $\tilde{\Phi}$ to distinguish effective potentials in each system of equations.
	
	\section{Jeans analysis for UFDs}
	
	The spherically symmetric Jeans equation when being projected along the line of sight yields the following dispersion velocity \cite{Binney2008}
	\begin{align}
	\sigma^2_p = \dfrac{2}{I(R)} \int_{R}^{\infty} \left( 1 - \beta \dfrac{R^2}{r^2} \right) \rho_*\sigma^2_r(r) \dfrac{rdr}{\sqrt{r^2 - R^2}}, \label{velo_disper}
	\end{align}
	where $I(R)$ is stellar surface density. From observations for UFDs, it is standard to fit this density with the Plummer profile given by \cite{Walker2009b}
	\begin{align}
	I(R) = \dfrac{L}{\pi R_h^2}\left( 1 + \dfrac{R^2}{R_h^2}\right)^{-2},
	\end{align}
	which deprojected in three-dimensional density of the form
	\begin{align}
	\rho_*(r) = \dfrac{3L}{4\pi R_h^3} \left( 1 + \dfrac{r^2}{R_h^2} \right)^{-5/2}.\label{de_stel_den}
	\end{align}
	Here, $R_h$ is called the half-light (effective) radius of the corresponding galaxy and its projected value can be related to the deprojected one by $r_{1/2} = (4/3) R_h$.
	\\
	
	The high mass-to-light ratio of UFDs identifies them as extreme cases for dark-matter dominated systems. In that case, stars are treated as "tracers" for an underlying gravitational potential generated by the corresponding mass distribution and their radial velocity dispersion can be calculated via \cite{Lokas:2003ks}
	\begin{align}
	\rho_* \sigma^2_r(r) = G r^{-2\beta}\int_{r}^{\infty} s^{2\beta - 2}\rho_*(s)M(s)ds. \label{rad_velo}
	\end{align}
	
	The velocity anisotropy $\beta$ is ambiguous for most of UFDs. However, a general radius-dependent profile can be taken into account as
	\begin{align}
	\beta(r) = (\beta_\infty - \beta_0)\dfrac{r^2}{r^2 + r^2_\beta} + \beta_0,
	\end{align}
	where $\beta_0$ and $\beta_\infty$ are asymptotic values at inner up to outer edge of the dark matter halo, $r_\beta$ is the scale radius of the stellar distribution. Because $\beta \equiv 1 - \sigma^2_r/\sigma^2_t < 1$ at all radius, we need to choose priors for $\beta_0, \beta_\infty$ satisfied this condition later on. \\
	
	In this work, we consider the dark matter profile with an inner soliton extending to NFW profile $\rho_{NFW}(r) = \rho_0 (r/r_s)^{-1}(1 + r/r_s)^{-2}$ at a transition radius of $r_{trans} = 3 r_c$ \cite{Marsh:2015wka}. The NFW characteristic density $\rho_0$ is determined by setting $\rho_s(r_{trans}) = \rho_{NFW}(r_{trans})$; $r_s$ characterizes the scale radius of NFW halos. The enclosed mass associated with this profile can be computed explicitly \cite{Chen:2016unw, Navarro:1995iw}
	\begin{multline}
	M(r) = M_s(r) = \dfrac{4.2 \times 10^7 \text{ M}_\odot}{m_{22}^2 (r / \text{pc})} \dfrac{1}{(a^2 + 1)^7} \left( 3465 a^{13} \right. \\ + 23100 a^{11}  + 65373 a^9 + 101376 a^7 + 92323 a^5 \\ \left. + 48580 a^3 -3465 a  + 3465(a^2 + 1)^7 \tan^{-1} (a) \right) \\ \text{ with } a = \left( 2^{1/8} - 1 \right)^{1/2} (r / r_c) \text{ when } r < r_{trans};
	\end{multline}
	\begin{multline}
	M(r) = M_s(r_{trans}) + M_{NFW}(r) - M_{NFW}(r_{trans}) \\
	= M_s(r_{trans}) + 4 \pi \rho_0 r^3_s \left[ \ln \left( \dfrac{r_s + r}{r_s + r_{trans}} \right) \right. \\ \left. - \dfrac{r}{r_s + r} + \dfrac{r_{trans}}{r_s + r_{trans}} \right] \text{when } r > r_{trans}.
	\end{multline}
	
	Notice that the tail of NFW profile makes the enclosed mass divergent as $\sim \ln r$ when $r \rightarrow \infty$. \\
	
	Therefore, to calculate the mass we integrate this profile up to a cut-off radius given by $r_{cut} = \text{Max}\left[r_{lim}, r_{Roche} \right]$. Here, $r_{lim}$ is about 10 times of the maximum stellar-extended radius; $r_{Roche}$ is the standard Roche limit for a point mass of $M_{halo} = 10^9 \text{ M}_\odot$ orbiting around Milky Way with a velocity dispersion of $\sigma_{MW} = 200 \text{ km s}^{-1}$,
	\begin{align}
	r_{Roche} \simeq \left( \dfrac{G M_h D^2}{2 \sigma_{MW}^2} \right)^{1/3},
	\end{align}
	and $D$ is the distance to the center of the dwarf satellite from the Earth. \\
	
	Following the same method in \cite{Strigari2008i, Wolf2010}, we have to determine a viable range for 6 free parameters referred as
	\begin{align}
	\vec{\theta} = \{m_{22}, r_c, r_s, r_\beta, \beta_0, \beta_\infty \}.
	\end{align}
	The likelihood function is chosen as a two-component Gaussian distribution 
	\begin{align}
	\mathcal{L}(\vec{\theta}) = \prod_{i=1}^N \dfrac{1}{\sqrt{2\pi \sigma^2_i}} \text{exp} \left[ -\dfrac{1}{2} \dfrac{(v_i - u)^2}{\sigma_i^2} \right].
	\end{align} 
	From observation, $v_i$ and $u$ are the heliocentric velocity for the $i^{th}$ star and the systemic velocity of the satellite, respectively. The velocity distribution is weighted by the total variance $\sigma^2_i = \sigma^2_{i, th} + \sigma^2_{i, m}$, where $\sigma^2_{i, th}$ is computed theoretically from \eqref{velo_disper} at a certain projected radius $R_i$ for an individual star while $\sigma^2_{i, m}$ refers to its velocity uncertainty in measurement. \\
	
	\begin{table}
		\footnotesize
		\centering
		\begin{tabular}{ c|c|c|c|c }
			UFDs & $R_h$ (pc) & $D$ (kpc) & $u$ (km s$^{-1}$) & Ref \\
			\hline 
			Segue 1 & 29 & 23 & 208.5 & \cite{Geha:2008zr, Simon:2010zr} \\
			Reticulum II & 55 & 32 & 62.8 & \cite{Simon:2015fdw} \\
			Carina II & 91 & 37.4 & 477.2 & \cite{Li:2018fdw} 
			\\
			Hydrus 1 & 53 & 27.6 & 80.4 & \cite{Koposov2018}
		\end{tabular}
		\caption{The structural parameters of UFDs and references to data sets used in this work.}
		\label{ufd_extrinsic}
	\end{table}
	
\begin{figure*}[!ht]
		\centering
		\includegraphics[scale=0.4]{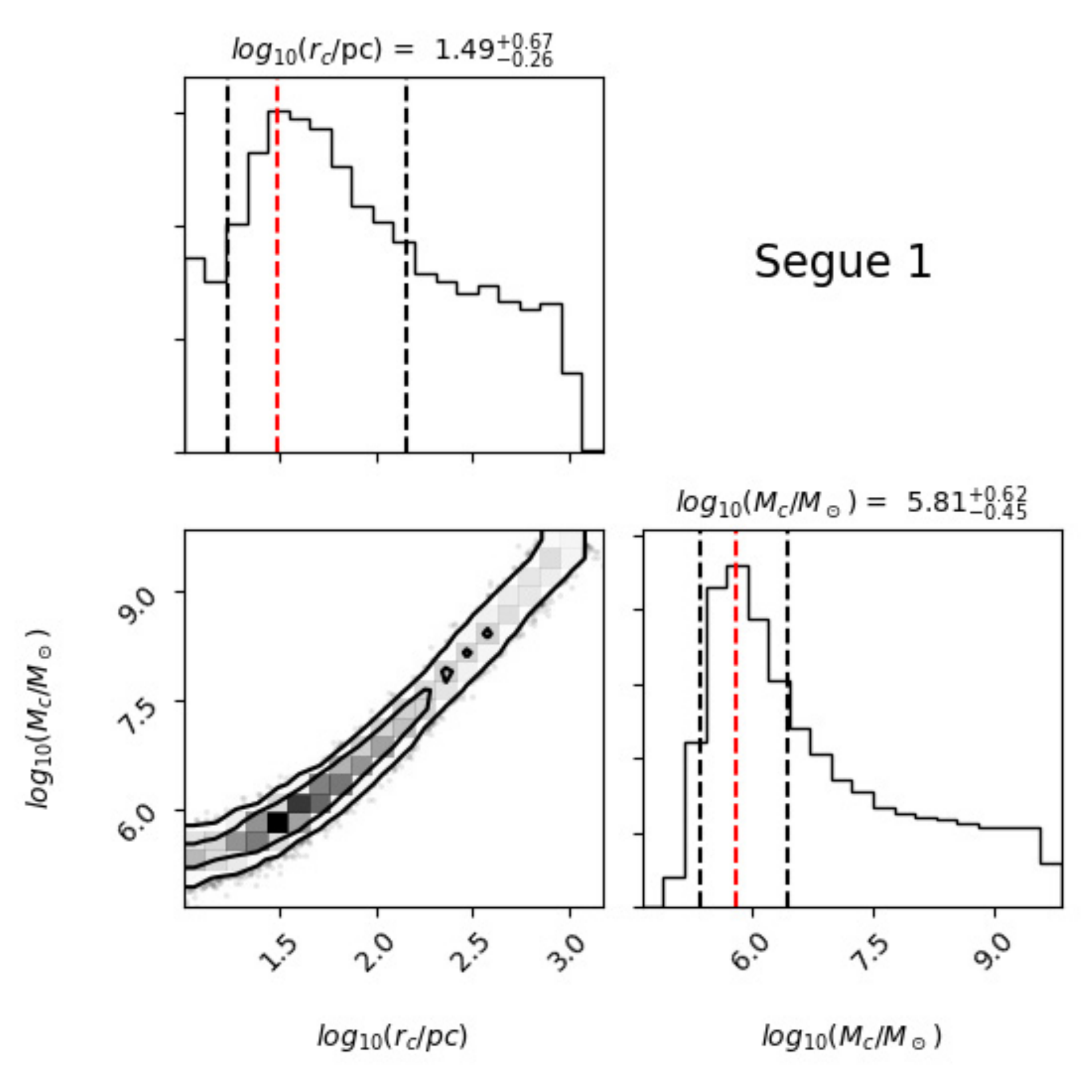}\includegraphics[scale=0.4]{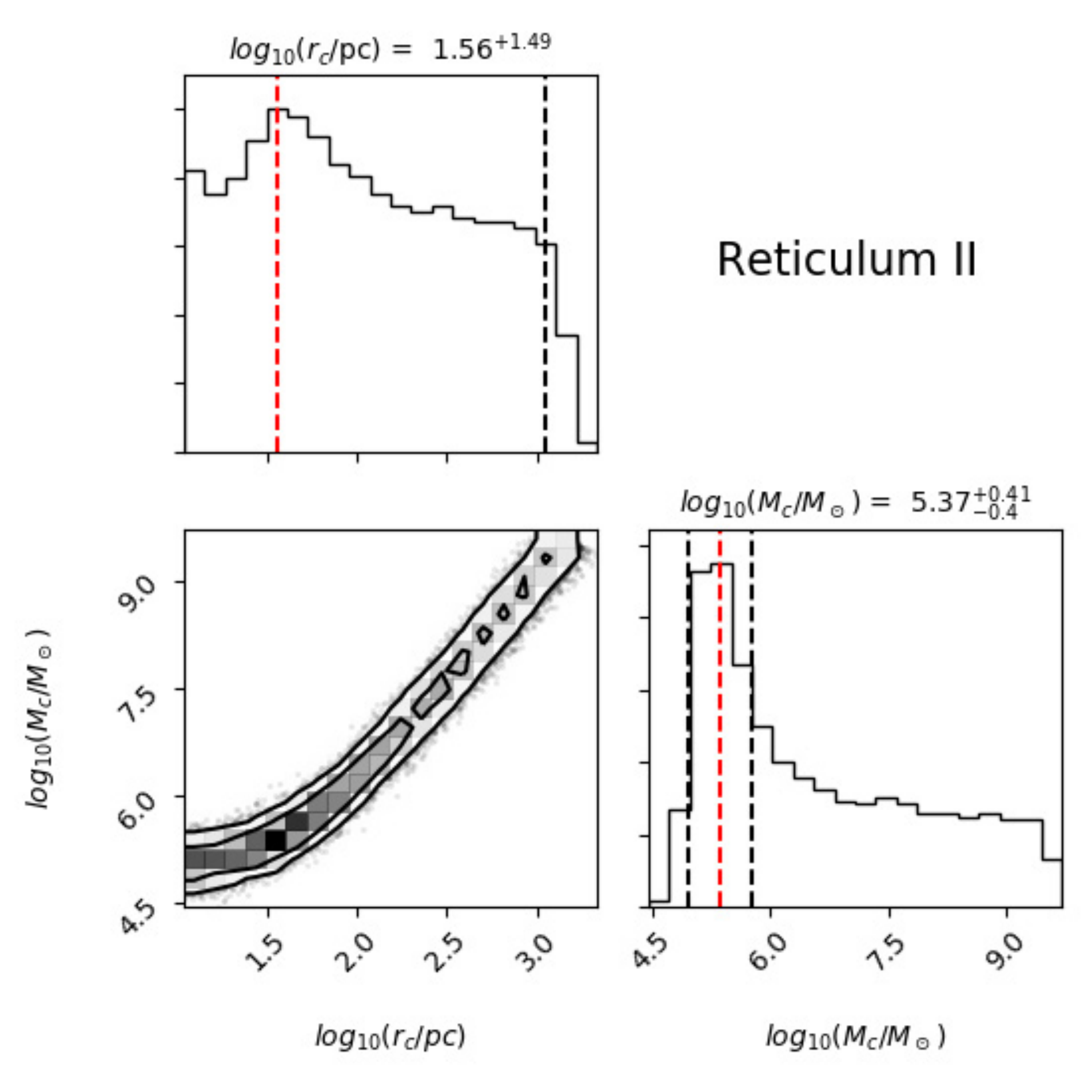} \\ \includegraphics[scale=0.4]{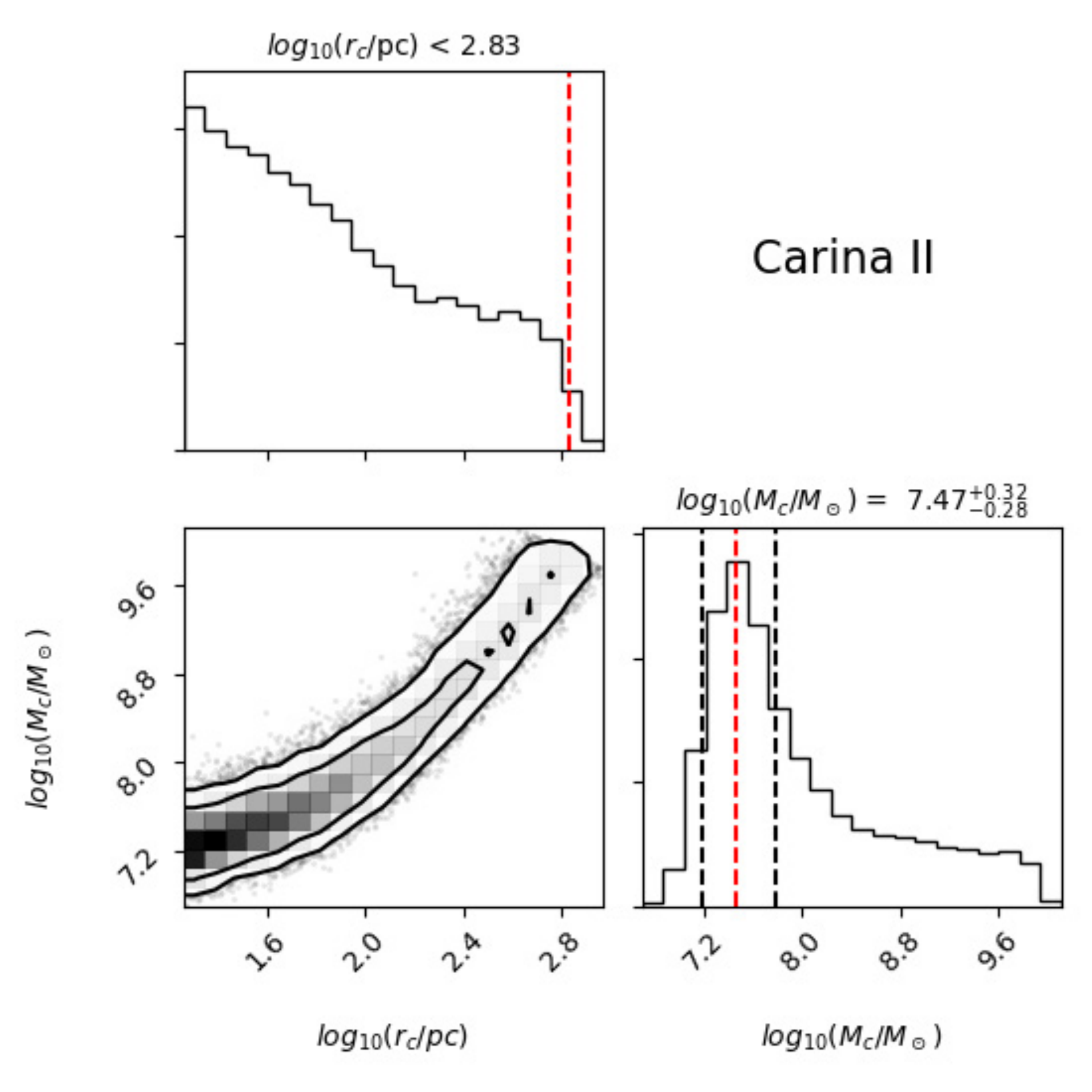}\includegraphics[scale=0.4]{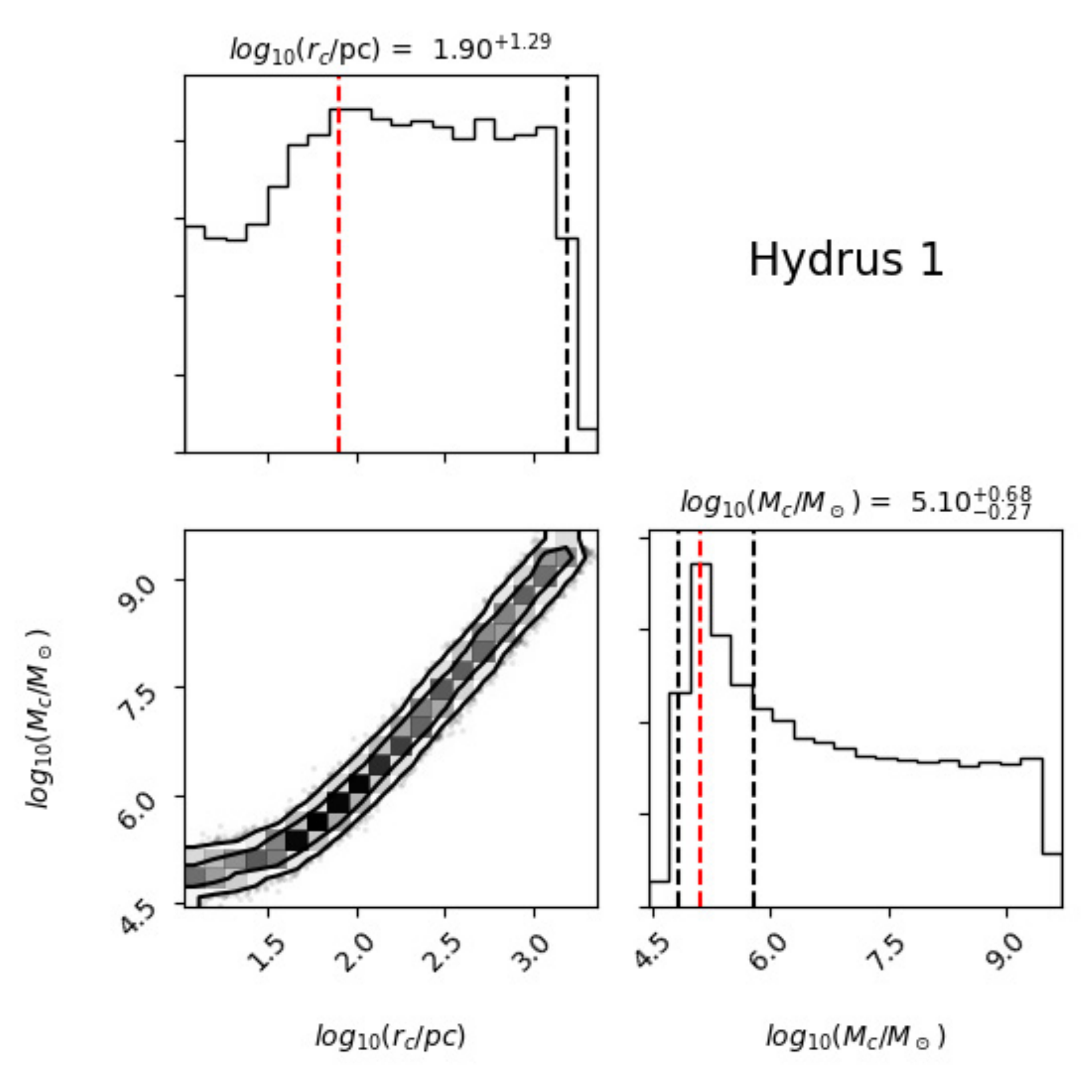}
		\caption{1D and 2D posterior distributions of four UFDs taken from MCMC chains using \textbf{emcee} \cite{Foremanmackey2012}, plotted using \textbf{corner} package \cite{corner2016}. Here, we simulate with 240 walkers and 10000 steps for each galaxy; only the last $3000 \times 240$ samples (thinning each 20 points) are chosen to plot. The red dashed-vertical lines in each 1D diagram show the peak value of the distributions and the black ones show errors at 60.6\% of the peak height. 68\% and 95\% confidence levels are shown by contours in the 2D correlation plots.}
		\label{ufds_samples}
	\end{figure*}

	Uniform priors are assumed on the parameters over the following ranges
	\begin{align}
	-1 < \log_{10}(m_{22}) < 5, &\quad 1 < \log_{10}(r_c/\text{pc}) < 4, \\
	1 < \log_{10}(r_s/\text{pc}) < 4, &\quad 1 < \log_{10}(r_\beta/\text{pc}) < 4, \\
	-10 < \beta_0 < 1, &\quad -10 < \beta_\infty < 1, \\
	0.2 R_h < &r_\beta < r_{lim}, \\
	0.2 R_h < r_c < 2 r_{cut}, &\quad 0.2 R_h < r_c < 2 r_{cut}.
	\end{align}
	In Table \ref{ufd_extrinsic}, we show properties of UFD candidates and references to their stellar data taken to analyze in this work. Most cases are confirmed as unlikely to be under ongoing tidal disruption; hence they are taken to be in dynamical equilibrium where Jeans analysis is applicable. Figures \ref{ufds_samples} illustrate distributions for two relevant quantities: $ M_c, r_c$ which can show their preferred values when being constrained from the data sets.
	
\end{appendices}

\end{document}